\documentclass{article}
\usepackage{bigsky07}
\usepackage{graphicx}
\usepackage{amssymb,amsmath}
\frompage{000} \topage{000}                                              
\def\rightmark{Analysis of Emission Shapes}
\def\leftmark{P.\ Danielewicz}

\title{Analysis of Emission Shapes}
\authors{
{Pawe\l\ Danielewicz$^{a}$ %
}\\[2.812mm]
{\normalsize
National Superconducting Cyclotron Laboratory and\\
Department of Physics and Astronomy, Michigan State University,\\
East Lansing, MI 48824, USA
}}

\abstract{
Shapes of relative emission sources can be accessed by expanding shapes of correlations at
low relative velocities in pair center of mass in Cartesian harmonics.
Coefficients of expansion for correlations are related to the respective
coefficients of expansion for the sources through one dimensional integral
transforms involving properties of pair relative wavefunctions.  The methodology
is illustrated with analyses of NA49 and PHENIX correlation data.
}


\begin{document}

\maketitle
\setcounter{page}{1}

\section{Introduction}\label{intro}

Correlations of particles at low relative velocities have been used to determine
sizes of emitting regions in heavy-ion reactions \cite{Wiedemann:1999qn,Lisa:2005}.
In general, the smaller the emitting region the stronger are final-state effects in emission and stronger correlations
between the emitted particles.  However, shapes of emitting regions are also of interest.
For example, a prolonged emission of particles, such as associated with a transitional behavior from quark-gluon plasma,
can produce a relative emission source elongated in the direction of pair total momentum \cite{rischkehydro}.
When the final-state effects are due to pure identity interference,
the correlation function represents a Fourier-transform
of the relative emission source \cite{Brown:1997ku}.  Most commonly,
in the past, source shapes have been analyzed for charged pions only, by representing the sources
in terms of a single anisotropic Gaussian, allowing to interpret
Coulomb-corrected correlation functions in terms of the Fourier-transformed Gaussian \cite{Wiedemann:1999qn,Lisa:2005}.
Here, I discuss an alternative strategy, where the correlation function and source functions are expanded in surface spherical harmonics.
Of the latter, the Carthesian harmonics~\cite{applequist89} exhibit particularly pleasing properties for the purpose.
The strategy is not confined to correlations dominated by identity interference.

\section{Correlations at Low Relative Velocity}

Possibility of learning on the geometry of emitting regions in reactions relies on the ability to
factorize the amplitude for the reaction into a wavefunction~$\Phi_{\bf q}^{(+)}$, for the pair of detected
particles, and an amplitude remnant.  At at low relative velocity, $|\Phi_{\bf q}^{(+)}|^2$
may exhibit pronounced spatial features such as associated with identity interference,
resonances or
Coulomb repulsion.  The wavefunction features can be regulated by changing the relative particle
momentum~${\bf q}$.  The amplitude
remnant, squared within a cross section and summed over the
unobserved particles and integrated over their momenta, yields $S'$ which
represents general features of reaction geometry, without a significant variation with~${\bf q}$.
When examining the inclusive two-particle cross-section then,
the structures in $|\Phi^{(+)}|^2$ can be used to explore the geometry in $S'$:
\begin{equation}
\frac{d \sigma}{d{\bf p}_1 \, d{\bf p}_2} =  \int  {\rm d} {\bf r} \,  S'_{\bf P} ({\bf r}) \,
 |\Phi^{(-)}_{{\bf q}} ( {\bf r})|^2 \, .
\end{equation}

The naive expectation, met at large $q$ for a multi-particle final state, is that the emission
of two particles is uncorrelated.  It is then interesting to normalize the two-particle cross section with
a product of single-particle cross sections and to look for the
evidence of correlations:
\begin{equation}
{\mathcal R}({\bf q}) =
\frac{\frac{1}{\sigma} \, \frac{d \sigma }{ d{\bf p}_1 \, d{\bf p}_2 } }
{\frac{1}{\sigma} \, \frac{d \sigma}{d{\bf p}_1 }   \, \frac{1}{\sigma}  \, \frac{d \sigma}{d{\bf p}_2 }}
- 1 = \int  {\rm d} {\bf r} \,
 \left( |\Phi^{(-)}_{{\bf q}} ( {\bf r})|^2 - 1 \right) \,
 S_{\bf P} ({\bf r}) \, .
 \label{eq:R}
\end{equation}
The second equality follows from the fact that, at large $q$, the relative wavefunction
squared is equal, on the average, to~1; in combining this with ${\mathcal R} \simeq 0$ at large $q$,
the source $S$, following
the cross-section normalization, turns out to be normalized to~1, $\int {\rm d} {\bf r} \, S({\bf r}) = 1$.
Equation (\ref{eq:R}) links the correlation function ${\mathcal R}$
to the interplay of deviations of the relative wavefunction from~1 with the geometry in~$S$.
Normalized to 1, $S$ may be interpreted as the probability density for emitting the two
particles at the relative separation~${\bf r}$ within the particle center of mass.
The emission is integrated over time, as far as $S$ is concerned.

\section{Correlations in Terms of Cartesian Harmonics}

Given ${\mathcal R}({\bf q})$ and $|\Phi|^2$, one can try to learn about
$S$; mathematically, this represents a difficult problem involving an
inversion of the integral kernel $K({\bf q}, {\bf r}) = |\Phi^{(-)}_{{\bf
q}} ( {\bf r})|^2 - 1$, in three dimensions.  The situation gets simplified by the fact
that $K$ depends only on $q$, $r$ and the angle $\theta_{\bf q \, r}$.  Correspondingly,
the kernel
can be expanded in Legendre polynomials:
\begin{equation}
K({\bf q},{\bf r}) =
\sum_\ell (2 \ell +1 )\, K_\ell (q,r) \, P^\ell
(\cos{\theta}) \, .
\end{equation}
Upon expanding the correlation and source functions in spherical
harmonics~$Y^{\ell m}$, ${\mathcal R} ({\bf q}) = \sqrt{4 \pi} \sum_{\ell
m} {\mathcal R}^{\ell m} (q) \, {\rm Y}^{\ell m} (\hat{\bf q})$, one finds that
the three-dimensional relation~(\ref{eq:R}) is equivalent to a set of one-dimensional
relations for the corresponding harmonic coefficients \cite{Brown:1997ku},
\begin{equation}
{\mathcal R}^{\ell m}(q)
= 4 \pi \int {\rm d}r \, r^2 \, K_\ell (q, r) \,
S^{\ell m} (r) \, .
 \label{eq:Rlm}
\end{equation}
For weak anisotropies, only low-$\ell$ coefficients of ${\mathcal R}$ or $S$ are expected to be
significant.  The $\ell=0$ version of (\ref{eq:Rlm}) connects the angle-averaged functions.

The above results suggest an analysis of the correlation functions and source
functions in
terms of the harmonic coefficients.  An issue, however, is that it is cumbersome to analyze
real functions in terms of complex coefficients that lack a clear interpretation for $\ell \ge 1$.
It becomes then natural to seek another basis for the directional decomposition of
correlation functions and sources, one that would be real and have a a clear geometric meaning.  Such a basis may
be constructed starting from a unit direction vector
$\hat{n}_\alpha = (\sin{ \theta} \cos{\phi}, \sin{ \theta} \sin{\phi}, \cos{ \theta})$.  A convenient choice of axes
relative to which the angles are determined in a reaction is $z$ along the beam axis, $x$ along the transverse component of pair total
momentum and $y$ perpendicular to the two other axes.

The tensor
product of $\ell$ vectors $\hat{n}_\alpha$ yields a symmetric rank-$\ell$ Cartesian tensor that is a
combination of spherical tensors of rank $\ell' \le \ell$ and the same evenness as~$\ell$:
\begin{equation}
(\hat{n}^{\ell})_{\alpha_1 \ldots \alpha_{\ell}} \equiv \hat{n}_{\alpha_1} \, \hat{n}_{\alpha_1} \ldots \hat{n}_{\alpha_\ell} =
\sum_{\ell' \le \ell, m} c_{\ell' m} \, {\rm Y}^{\ell' m} \, .
\end{equation}
A projection operator $\mathcal{P}$ may be constructed in the space of symmetric
rank-$\ell$ tensors, out of a combination of Kronecker $\delta$-symbols, that makes a symmetric
tensor traceless,
\begin{equation}
\sum_{\alpha} ({\mathcal P} \hat{n}^\ell)_{\alpha \, \alpha \, \alpha_3 \ldots \alpha_\ell} = 0
\, .
\end{equation}
The tracelessness of $({\mathcal P} \hat{n}^\ell)$ ensures \cite{applequist89,danielewicz-2007-75} that the products
$r^\ell \, ({\mathcal P} \hat{n}^\ell)_{\alpha_1 \ldots \alpha_\ell}$
are solutions of the Laplace equation and, thus, the
$({\mathcal P} \hat{n}^\ell)_{\alpha_1 \ldots \alpha_\ell} \equiv {\mathcal A}_{\alpha_1 \ldots \alpha_\ell}^{(\ell)} \equiv
{\mathcal A}_{x^{\ell_x} \, y^{\ell_y} \, z^{\ell_z} }^{(\ell)}$ are combinations of spherical
harmonics of rank $\ell$ only.  In the above, $\ell_x$, $\ell_y$ and~$\ell_z$ are the number of times of repeated, respectively, $x$, $y$ and $z$
indices in the symmetric tensor ${\mathcal P} \hat{n}^\ell \equiv {\mathcal A}^{(\ell)}$, $\ell_x + \ell_y + \ell_z = \ell $.
The Cartesian tensor components of
$ {\mathcal A}^{(\ell)} $ are real and may be used to
replace~$Y^{\ell m}$.  The lowest-rank tensors are:
\begin{eqnarray}
{\mathcal A}^{(0)}  =  1 \, , \hspace*{1.5em}
{\mathcal A}_\alpha^{(1)})  =  \hat{n}_\alpha \, , \hspace*{1.5em}
{\mathcal A}_{\alpha_1 \, \alpha_2}^{(2)})  =   \hat{n}_{\alpha_1} \,  \hat{n}_{\alpha_2}
- \frac{1}{3} \delta_{\alpha_1 \, \alpha_2} \, , \nonumber \\
{\mathcal A}_{\alpha_1 \,
\alpha_2 \, \alpha_3}^{(3)}  =  \hat{n}_{\alpha_1} \,  \hat{n}_{\alpha_2} \,  \hat{n}_{\alpha_3}
- \frac{1}{5} ( \delta_{\alpha_1 \, \alpha_2} \, \hat{n}_{\alpha_3}
+ \delta_{\alpha_1 \, \alpha_3} \, \hat{n}_{\alpha_2} + \delta_{\alpha_2 \, \alpha_3} \, \hat{n}_{\alpha_1}) \, .
\end{eqnarray}

The completeness relation in terms of the Cartesian components is \cite{danielewicz-2007-75}
\begin{eqnarray}
\delta(\Omega' - \Omega) & =  & \frac{1}{4 \pi} \sum_{\ell} \frac{(2\ell+1)!!}{\ell !}
\sum_{\alpha_1 \ldots \alpha_\ell}
{\mathcal A} _{\alpha_1 \ldots \alpha_{\ell}}^{(\ell)} \,
{\mathcal A} _{\alpha_1 \ldots \alpha_{\ell}}^{(\ell)} \nonumber \\
& = & \frac{1}{4 \pi} \sum_{\ell} \frac{(2\ell+1)!!}{\ell !}
\sum_{\alpha_1 \ldots \alpha_\ell}
{\mathcal A}  _{\alpha_1 \ldots \alpha_{\ell}}^{(\ell)} \,
\hat{n}_{\alpha_1} \ldots \hat{ n}_{\alpha_{\ell}} \, ,
\end{eqnarray}
where the second equality follows from ${\mathcal D}={\mathcal D}^\top={\mathcal D}^2$.
The completeness relation can be utilized for expanding ${\mathcal R}$ or $S$ in terms of
the Cartesian tensor components
\begin{equation}
\label{eq:Rexp}
{\mathcal R}({\bf q})= \int {\rm d}\Omega' \, \delta(\Omega' - \Omega) \, {\mathcal R}({\bf q}') = \sum_\ell \sum_{\alpha_1 \ldots \alpha_\ell}
{\mathcal R}_{\alpha_1 \ldots \alpha_\ell}^{(\ell)}(q) \, \hat{ q}_{\alpha_1} \ldots \hat{ q}_{\alpha_\ell}
\, ,
\end{equation}
where the coefficients are angular moments,
\begin{equation}
\label{eq:Rl}
{\mathcal R}_{\alpha_1 \ldots \alpha_\ell}^{(\ell)} (q) =
\frac{(2\ell+1)!!}{\ell !} \int \frac {{\rm d} \Omega_{\bf q}}{4 \pi} \,
{\mathcal R}({\bf q}) \, {\mathcal A} _{\alpha_1 \ldots \alpha_{\ell}}^{(\ell)} \, .
\end{equation}

With (\ref{eq:Rlm}) and with
the ${\mathcal R}$ and $S$ Cartesian coefficients being identical combinations
of the respective $Y^{\ell m}$-coefficients, the corresponding ${\mathcal R}$ and $S$
Cartesian coefficients are directly related to each other,
\begin{equation}
\label{eq:RKS}
{\mathcal R}_{\alpha_1\cdots\alpha_\ell}^{(\ell)}(q)
= 4 \pi \int  {\rm d}r \, r^2 \,
K_\ell(q,r) \,
{\mathcal S}_{\alpha_1\cdots\alpha_\ell}^{(\ell)}(r) \, .
\end{equation}
For weak anisotropies, only low-$\ell$ Cartesian coefficients for the source or correlation matter,
\begin{equation}
\label{eq:Rq}
{\mathcal R}({\bf q})  =
{\mathcal R}^{(0)}(q) + \sum_\alpha \, {\mathcal R}^{(1)}_\alpha(q) \,
\hat{q}_\alpha + \sum_{\alpha_1 \, \alpha_2}
{\mathcal R}^{(2)}_{\alpha_1 \alpha_2} (q) \,
\hat{q}_{\alpha_1} \, \hat{q}_{\alpha_2} + \ldots
\, .
\end{equation}

The correlation coefficients ${\mathcal R}^{(\ell)} $ allow to summarize three-dimensional information accumulated in the measurements,
in terms of one-dimensional plots.  With~(\ref{eq:Rexp}) and (\ref{eq:Rq}), the values of correlation (and similarly source) function
in the $+z$ direction are given by
\begin{equation}
\nonumber
{\mathcal R}(q) = {\mathcal R}^{(0)}(q) + {\mathcal R}_z^{(1)}(q) + {\mathcal R}_{zz}^{(2)}(q) + \ldots
\end{equation}
and in the $xz$ plane at 45$^\circ$ degrees by
\begin{equation}
\nonumber
{\mathcal R}(q) = {\mathcal R}^{(0)}(q) + \frac{1}{\sqrt{2}}{\mathcal R}_z^{(1)}(q)
+ {\mathcal R}_{xz}^{(2)}(q)
+ \frac{1}{2} \left( {\mathcal R}_{xx}^{(2)}(q)
+ {\mathcal R}_{zz}^{(2)}(q) \right)
+ \ldots \, .
\end{equation}
The coefficients ${\mathcal R}^{(\ell)}$ can be found through angular integration (\ref{eq:Rl}) or by fitting the expansion series
(\ref{eq:Rq}) to the dependence of data on spherical angle.
The latter procedure can be particularly useful for an experiment blind in certain directions.  For relating the correlation coefficients
to source coefficients, the two-body problem needs to be solved for the wavefunction $\Phi^{(-)}_{{\bf q}} ( {\bf r})$.  Not only
identity interference, but also Coulomb and strong interactions give rise to structures in $|\Phi^{(-)}_{{\bf q}} ( {\bf r})|^2$ that yield
finite kernels $K_\ell$ for multipolarities $\ell > 0$ \cite{danielewicz-2007-75}.  Contrary to naive expectations,
even strong interactions acting in $s$-wave only produce
finite kernels $K_\ell$ at $\ell > 0$.  With regard to the Coulomb interactions, there is no good reason trying to correct for them
\cite{Lisa:2005}, as they can provide access to source deformations just as the identity interference can.
Features of the source can be accessed in (\ref{eq:RKS}) by assuming a parameterized shape of the source and fitting source parameters to the correlation
coefficients, or by imaging source coefficients in the same manner as for $\ell=0$
\cite{Brown:2000aj}.

\section{Sample Results from Data Analysis}

The number of angular expansion coefficients that need to be considered for the correlation and source functions can be quite small.
Thus, for pairs of identical particles, such as identical pions, the correlation functions are symmetric
under inversion of relative momentum, ${\pmb q} \rightarrow -{\pmb q}$, meaning that only
even-$\ell$ components are finite.  Around midrapidity of a symmetric system, such as Pb + Pb or Au + Au,
the correlation function should be symmetric
with respect to the interchange of forward and backward directions, $q_z \rightarrow - q_z $, meaning that
only even-$z$ components of the functions can be finite.  When the correlation functions are averaged over
reaction-plane orientation, they are symmetric under interchange of the orientation of sideway axis,
$q_y \rightarrow - q_y $, meaning that only even-$y$ components of the functions can be finite.
In the end, as a consequence of the above, the correlation function in the considered situation must be invariant under the interchange
of the orientation of outward axis, $q_x \rightarrow - q_x $, meaning that only even-$x$ components of the functions can be finite.
For $\ell \le 4$, the finite Cartesian moments are then $\ell=0$, $x^2$, $y^2$ and $z^2$ for $\ell=2$ and $x^4$, $y^4$, $z^4$,
$x^2 \, y^2$, $x^2 \, z^2$ and $y^2 \, z^2$ for $\ell=4$.  Due to the tracelessness of the tensors from Cartesian components,
however, only 2 of the above $\ell=2$ components are independent and only 3 of the $\ell=4$.  Thus, depending on data statistics,
the dependence of anisotropic correlation function on spherical angle at any $q$ can be described with either 3,
for $\ell \le 4$, or 6, for $\ell \le 6$, functions
of relative momentum~$q$.

\begin{figure}[htb]
\begin{center}
\parbox{2.45in}{
\includegraphics[width=1.02\linewidth,height=3.12in]{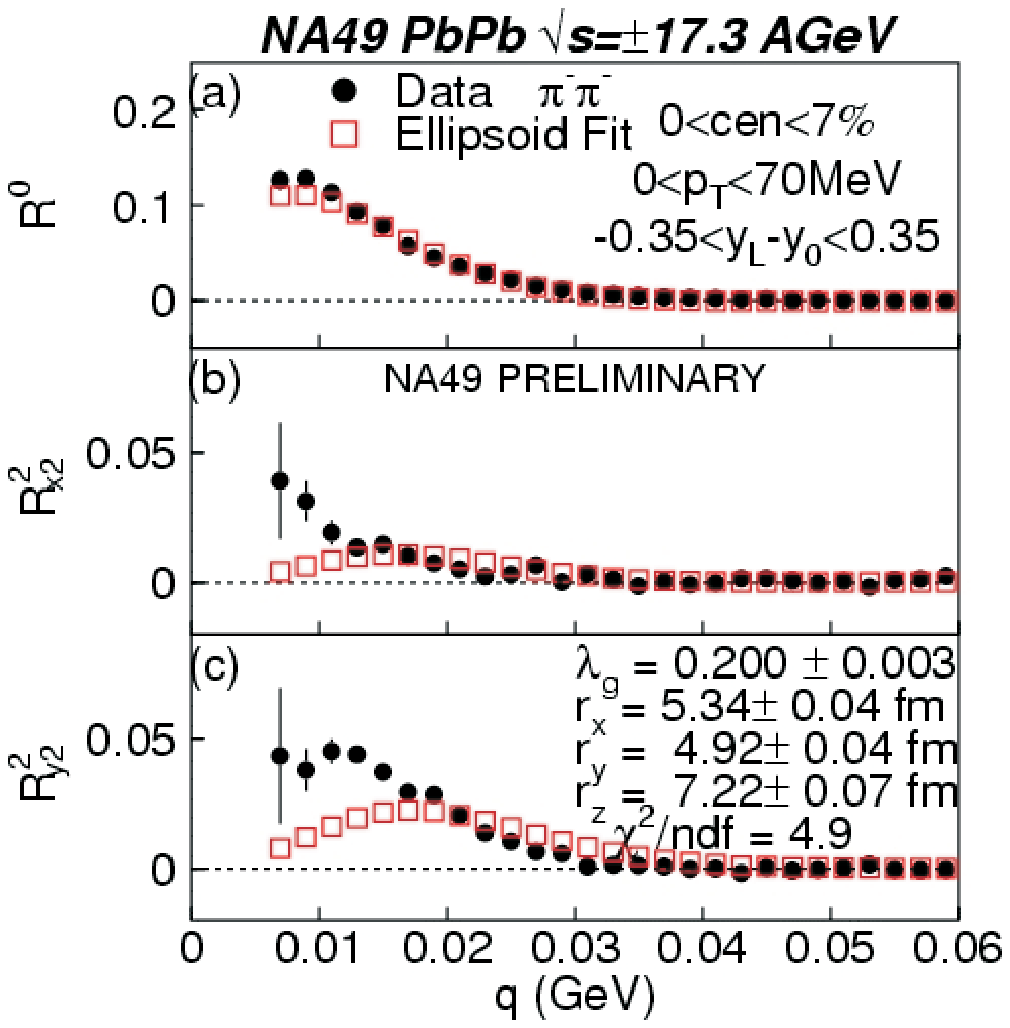}
}
\parbox{2.45in}{
\includegraphics[width=1.02\linewidth,height=3.12in]{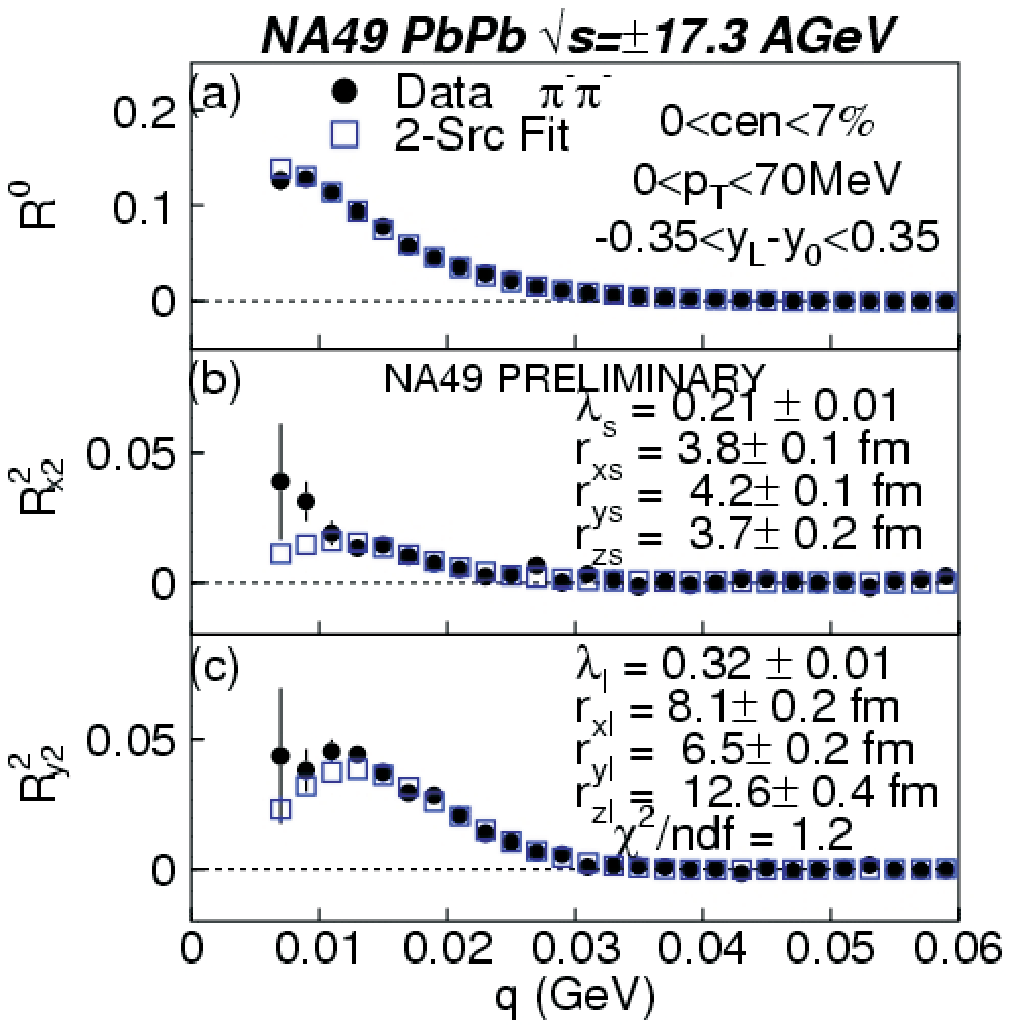}
}
\end{center}
\vspace*{-.6cm}
\caption[]{$\ell = 0$ and two independent $\ell=2$ angular moments of $\pi^-$-$\pi^-$ correlation functions
in central Pb + Pb collisions at $\sqrt{s}=17.3$~A\,GeV.  Filled circles represent measurements of the
NA49 Collaboration  \cite{Chung:2006}.  Squares in the left panels represent a fit to the data with
a source in the form of an anisotropic Gaussian.  Squares in the right panels represent a fit to the
data with a source combining two anisotropic Gaussians.
}
\label{fig:49Corr}
\end{figure}

Figure \ref{fig:49Corr} shows $\ell=0$ and $\ell=2$ Cartesian harmonic results for the midrapidity $\pi^-$-$\pi^-$ correlation functions
measured by the NA49 Collaboration in central Pb + Pb collisions at $\sqrt{s}=17.3$~A\,GeV.  The $\ell=0$ panels show
the correlation function averaged over directions of relative momentum and the $\ell=2$ panels illustrate the
quadrupole deformation of the function.  The $z^2$ coefficients are given by negative of the sum of $x^2$ and $y^2$
coefficients, ${\mathcal R}_{z^2}^{(2)} = - ({\mathcal R}_{x^2}^{(2)} + {\mathcal R}_{y^2}^{(2)})$.
No Coulomb corrections have been applied to the data.  On the other hand,
effects of Coulomb interactions are accounted for directly in relating source functions to the data.  In the left panels
of Fig.\ \ref{fig:49Corr}, it is seen that a single Gaussian source fails to describe
the data at low relative momenta.  In the right panels, on the other hand, it is seen that a source combining two Gaussian
sources provides a much better description of the data.

\begin{figure}[htb]
\begin{center}
\parbox{2.45in}{
\includegraphics[width=1.02\linewidth,height=3.12in]{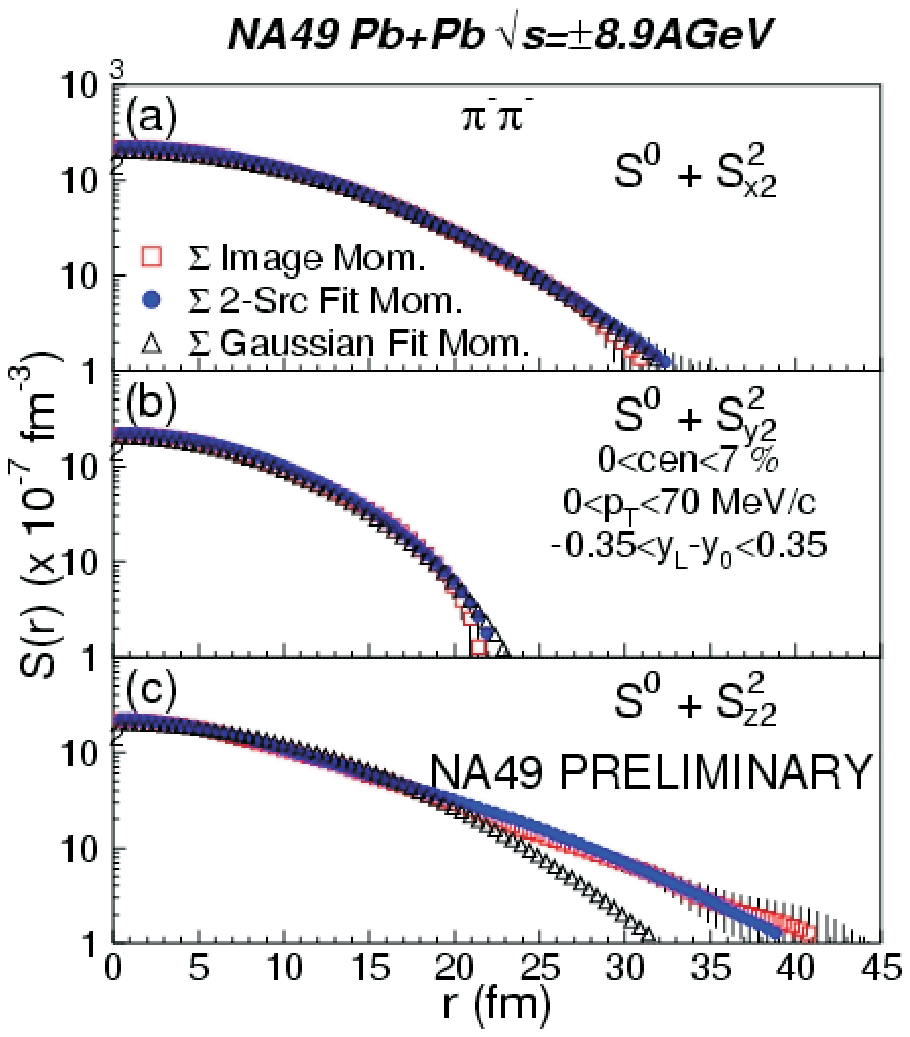}
}
\parbox{2.45in}{
\includegraphics[width=1.02\linewidth,height=3.12in]{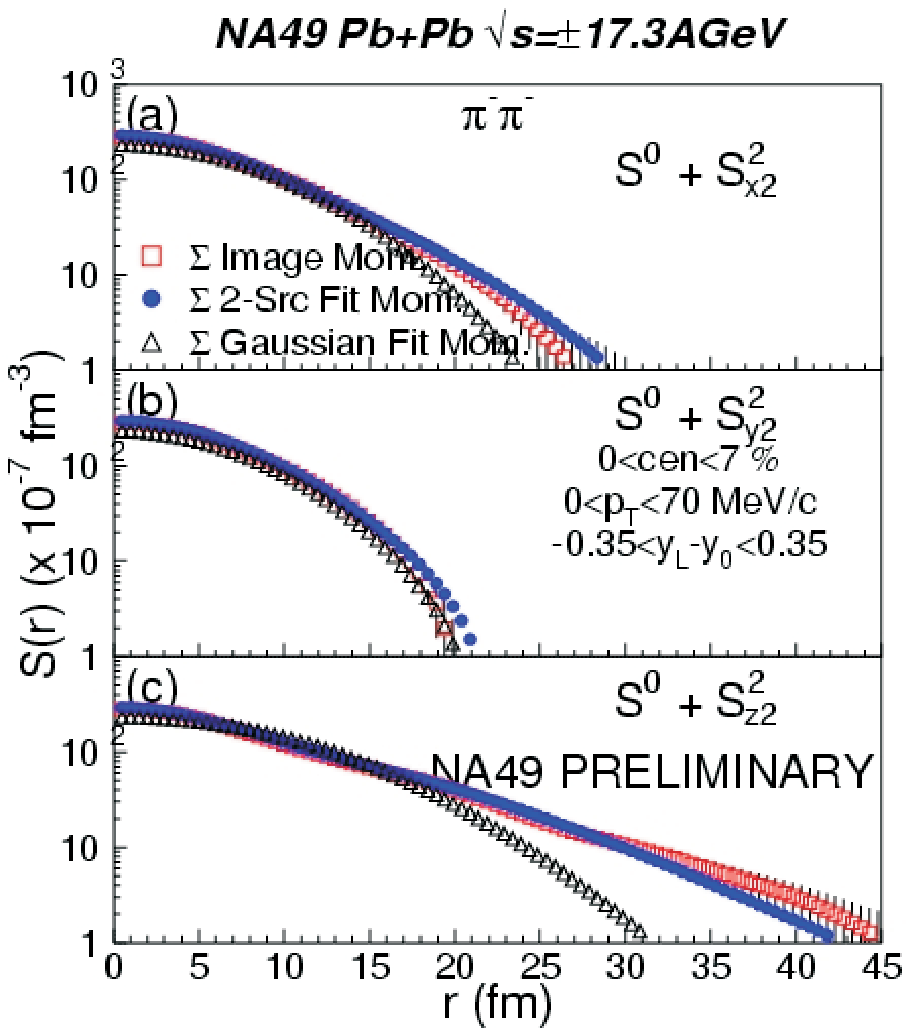}
}
\end{center}
\vspace*{-.6cm}
\caption[]{Relative midrapidity $\pi^-$-$\pi^-$ source in central Pb + Pb collisions
along $x$, $y$ and $z$ coordinate axes, from summing the $\ell \le 2$ angular components of the source.
The left and right panels represent, respectively, source in collisions at $\sqrt{s}=8.9$~A\,GeV and
$\sqrt{s}=17.3$~A\,GeV.  Squares represent
source values obtained by imaging the angular components of correlation functions measured by the NA49
Collaboration \cite{Chung:2006}.
Triangles represent results of the fit to the measured correlation moments \cite{Chung:2005} with a source in the form of an anisotropic
Gaussian.  Circles represent results of the fit to the measured correlation moments with a source combining two anisotropic Gaussians.
}
\label{fig:49Source}
\end{figure}

Figure \ref{fig:49Source} shows next source values along $x$, $y$ and $z$ axes, from summing $\ell = 0$ and
$\ell=2$ source coefficients deduced from the coefficients of correlation function measured in central
Pb + Pb collisions at $\sqrt{s}=8.9$~A\,GeV and
$\sqrt{s}=17.3$~A\,GeV \cite{Chung:2006}.  Aside from the single and two Gaussian results, also source values
obtained through imaging \cite{Brown:2000aj} are shown.  The imaged sources usually yield a very good description
of measured correlations.  It is seen in the figured that the fitted 2-Gaussian and imaged sources agree fairly well
with each other.  At both energies, those sources exhibit tails extending over 30~fm of relative distance, in the longitudinal $z$-direction.
The failure of the single-Gaussian source to reproduce the correlation moments at low~$q$, observed in Fig.~\ref{fig:49Corr}, appears to be related
to the failure in producing such a tail at large~$r$.  A non-Gaussian enhancement in the outward $x$-direction further develops
in the imaged and 2-Gaussian sources between the energies of $\sqrt{s}=8.9$~A\,GeV and
$\sqrt{s}=17.3$~A\,GeV. The sources in Fig.~\ref{fig:49Source} are most compact in the $y$ direction.
A tail in the source in the $z$ direction is expected in connection with a strong collective
expansion of the system along the beam \cite{danielewicz-2007-75}.  Greater extension in the $x$- than the $y$- direction and the tail in the $x$-direction
can be associated with
emission extended over time.  Low transverse momenta have been purposely selected for the analysis to eliminate any significant effects of
Lorentz $\gamma$-factors.

\begin{figure}[htb]
\begin{center}
\parbox{2.45in}{
\includegraphics[width=1.02\linewidth,height=3.12in]{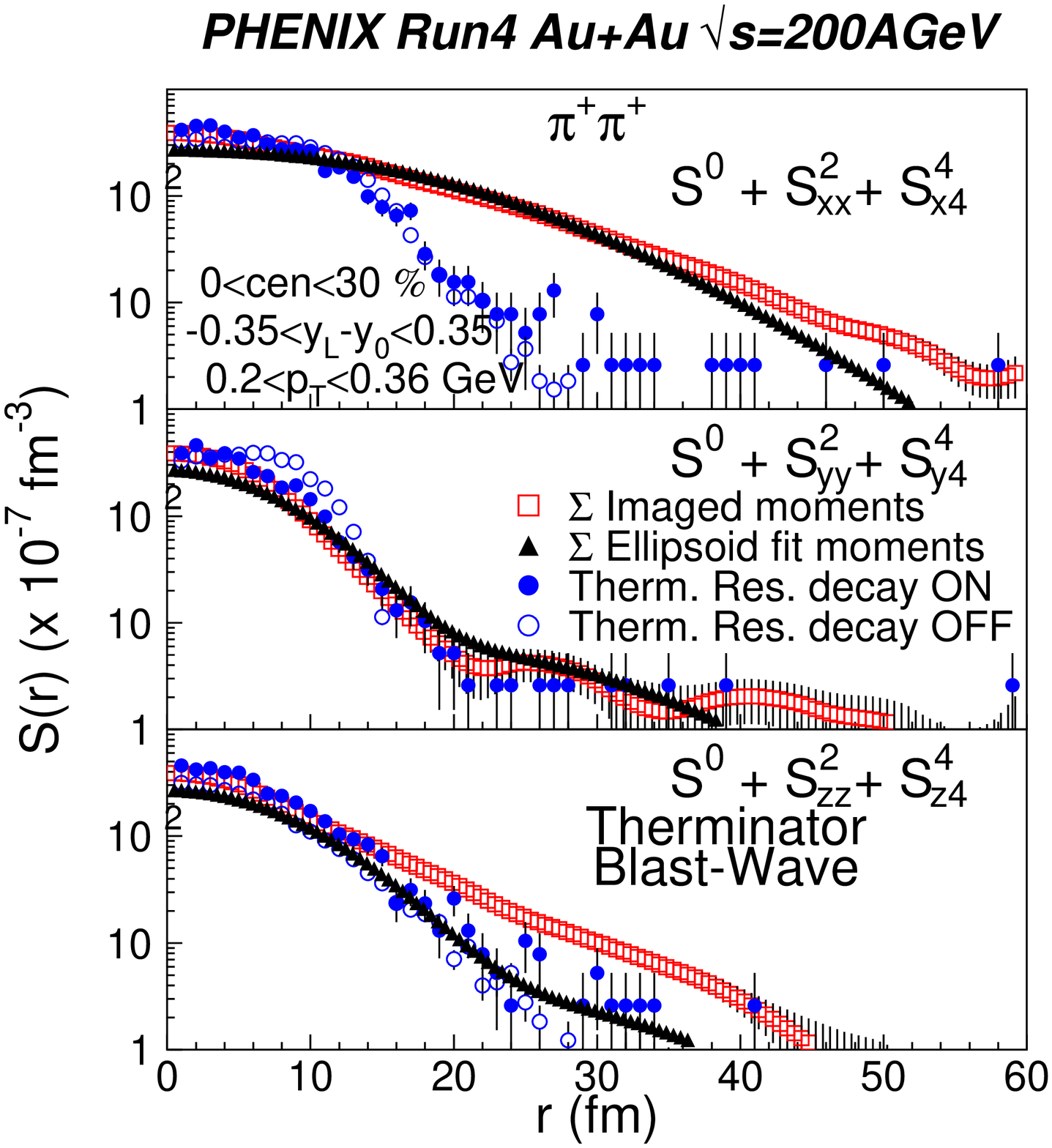}
}
\parbox{2.45in}{
\includegraphics[width=1.02\linewidth,height=3.12in]{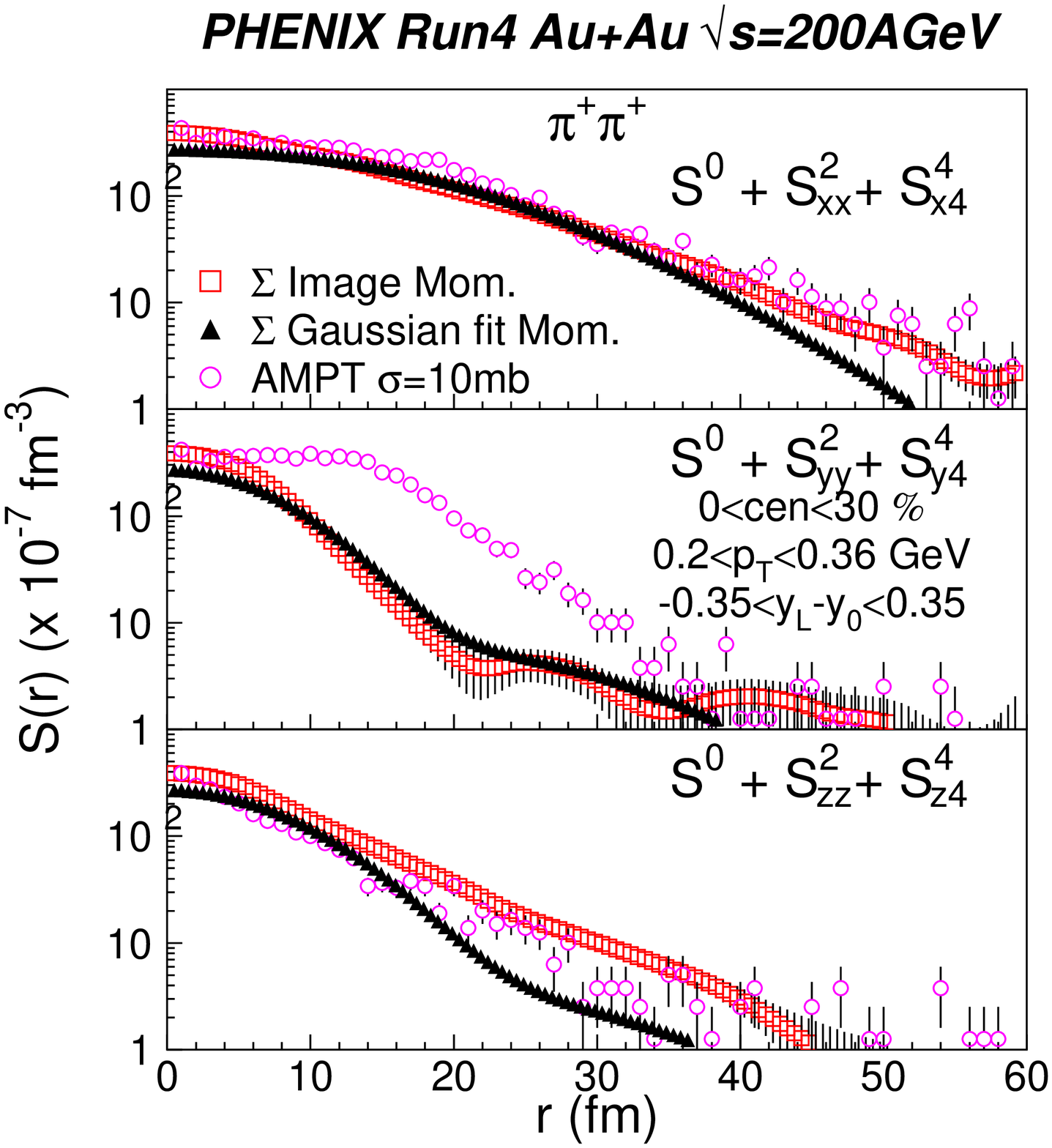}
}
\end{center}
\vspace*{-.6cm}
\caption[]{Relative midrapidity $\pi^+$-$\pi^+$ source in central Au + Au collisions at
$\sqrt{s}=200$~A\,GeV, along $x$, $y$ and $z$ coordinate axes, from summing the $\ell \le 4$ angular components of the source.  Squares represent
source values obtained by imaging the angular components of correlation functions measured by the PHENIX
Collaboration \cite{Chung:2005}.
Triangles represent results of the fit to the measured correlation moments \cite{Chung:2005} with a source in the form of an anisotropic
Gaussian.  In the left panels, circles represent a fit using
the Therminator model \cite{kisiel-2006-73,kisiel-2006-174} with, respectively, resonance
decays operating, for full circles, and switched off, for open circles.  In the right panels, circles represent results of the
AMPT model~\cite{Zhang:1999bd}.}
\label{fig:PhSource}
\end{figure}

Extended non-Gaussian tails in the longitudinal $z$-direction are further observed in central collisions at RHIC energies.
Figure \ref{fig:PhSource} shows the source values extracted from the analysis of $\ell \le 4$ angular moments of midrapidity correlations
measured by the PHENIX Collaboration in central Au + Au collisions at $\sqrt{s}=200$~A\,GeV.  Again, a fit with a single Gaussian
to the correlation functions fails to produce the tail along $z$-axis emerging from the imaging.  The imaged source also exhibits
some tail in the outward direction which is not there in the single Gaussian source.

Figure \ref{fig:PhSource} compares further the sources extracted from PHENIX data to those predicted by model calculations.
The Therminator model \cite{kisiel-2006-73}, represented in the left panels, has been successful in the past to reproduce
Gaussian-source parameters extracted from correlation sources corrected for Coulomb effects.  The model supplements short-range
features of a source with long-range features due to resonance decays.  In the left panels of Fig.~\ref{fig:PhSource}, it is observed
that the model can fully explain only the experimental source values in the $y$-direction and partly in the $z$-direction.  However, the model
fails completely as far as the long-range features of the imaged source in the $x$ direction.  The latter is quite independent of whether
the resonance decays take place or are suppressed.

The multi-phase transport model (AMPT) \cite{Zhang:1999bd} follows the dynamics of both partonic and hadronic phases and includes
resonance decays.  As is apparent in the right panels of Fig.~\ref{fig:PhSource}, the model describes fairly well the imaged source features
along the $x$ and $z$ directions but fails rather dramatically with respect to the $y$ direction.

\section{Conclusions}\label{concl}
As demonstrated with examples of the analyses of NA49 and PHENIX correlation measurements, the expansion in Cartesian harmonics
represents a practical strategy for summarizing correlation functions and for accessing sources shapes.  The strategy is capable of
producing a wealth of detail on emission sources and challenging this way theoretical interpretations of the data.

\section*{Acknowledgments}
Results in this paper stem from collaborations with David Brown, Paul Chung and Scott Pratt.  This work was supported
by the U.S.\ National Science
Foundation under Grants PHY-0245009 and PHY-0555893.

\section*{Note}
\begin{notes}
\item[a]
E-mail: danielewicz@nscl.msu.edu
\end{notes}

\bibliographystyle{bigsky07}
\bibliography{cobs07}

\begin{thebibliography}{10}
\expandafter\ifx\csname url\endcsname\relax
  \def\url#1{{\tt #1}}\fi
\expandafter\ifx\csname urlprefix\endcsname\relax\def\urlprefix{URL }\fi

\bibitem{Wiedemann:1999qn}
U.~A. Wiedemann and U.~W. Heinz, {\it Phys. Rept.\/} {\bf 319} (1999) 145.

\bibitem{Lisa:2005}
M.~Lisa, S.~Pratt, R.~Soltz and U.~Wiedemann, {\it Annu. Rev. Nucl. Part.
  Sci.\/} {\bf 55} (2005) 357.

\bibitem{rischkehydro}
M.~Gyulassy and D.~Rischke, {\it Nucl. Phys. A\/} {\bf 608} (1996) 479.

\bibitem{Brown:1997ku}
D.~A. Brown and P.~Danielewicz, {\it Phys. Lett.\/} {\bf B398} (1997) 252.

\bibitem{applequist89}
J.~Applequist, {\it J. Phys. A: Math. Gen.\/} {\bf 22} (1989) 4303.

\bibitem{danielewicz-2007-75}
P.~Danielewicz and S.~Pratt, {\it Phys. Rev. C\/} {\bf 75} (2007) 034907.

\bibitem{Brown:2000aj}
D.~A. Brown and P.~Danielewicz, {\it Phys. Rev. C\/} {\bf 64} (2001) 014902.

\bibitem{Chung:2006}
P.~Chung and P.~Danielewicz, in {\it Proceedings of Workshop on Particle
  Correlations and Femtoscopy, Sao Paulo, 2006\/}  .

\bibitem{Chung:2005}
P.~Chung, P.~Danielewicz, W.~Holzmann, R.~Lacey and J.~Alexander, in {\it
  Proceedings of Workshop on Particle Correlations and Femtoscopy, Kromeriz,
  2005\/}  .

\bibitem{kisiel-2006-73}
A.~Kisiel, W.~Florkowski, W.~Broniowski and J.~Pluta, {\it Phys. Rev. C\/} {\bf
  73} (2006) 064902.

\bibitem{Zhang:1999bd}
B.~Zhang, C.~M. Ko, B.-A. Li and Z.-W. Lin, {\it Phys. Rev. C\/} {\bf 61}
  (2000) 067901.

\bibitem{kisiel-2006-174}
A.~Kisiel, T.~Taluc, W.~Broniowski and W.~Florkowski, {\it Computer Physics
  Communications\/} {\bf 174} (2006) 669.

\end{thebibliography}

\vfill\eject
\end{document}